\documentclass[aps,twocolumn,showpacs,floats,floatfix,prl]{revtex4}

\usepackage[T1]{fontenc}
\usepackage[latin9]{inputenc}
\usepackage{times}
\usepackage{graphicx}
\usepackage{dcolumn}
\usepackage{bm}
\usepackage{color}
\usepackage{amsmath}
\usepackage{subfigure}
\usepackage{enumitem}
\usepackage{epstopdf}
\usepackage{hyperref}
\usepackage{xcolor}
\usepackage{ulem}
\usepackage{footnote}
\usepackage{hyperref}

\hypersetup{
    colorlinks=true,
    linkcolor=blue,
    filecolor=magenta,      
    urlcolor=purple,
    citecolor=cyan
}

\begin{document}

\author{L. A. Pe\~{n}a Ardila}

\affiliation{Institut f\"{u}r Theoretische Physik, Leibniz Universit\"{a}t, 30167 Hannover, Germany}

 \newcommand{\LA}[1]{{\color{red}#1}}

\title{Dynamical formation of polarons in a Bose-Einstein condensate: A variational approach}

\begin{abstract} 
We investigate the non-equilibrium dynamics of an impurity coupled to a Bose-Einstein condensate,  systematically compared  with recent experimental results [M. G. Skou et al., Nat. Phys. (2021)]. The dynamics of the impurity is tracked down by using a time-dependent variational coherent ansatz. For weak coupling between the impurity and the bath, analytical expressions for the time-dependent contrast are derived, matching quite well with previous findings obtained within a master equation approach. For strong coupling instead, the variational ansatz provides a good quantitative description of the polaron dynamics, in particular, in signaling the transition from the few to the many-body correlated regime where polarons are expected to form. 
\end{abstract}

\maketitle

\section{I. Introduction}
\label{sec:intro}

One of the most challenging problems in current quantum science and technology has to do with providing a complete description of a many-body system. Besides the large number of degrees of freedom, interactions and correlations make the problem even harder to  tackle. The concept of quasiparticle arises as an alternative for mapping this complex problem into a more tractable one. In fact, quasiparticles are crucial for understanding several phenomena ranging from atomic and nuclear  to high-energy physics. One of the most famous paradigms of quasiparticles relies on Pekar and Landau's original proposal of the polaron concept~\cite{Landau48}. When an electron travels through a material, it gets dressed by the low-energy excitations of the lattice (phonons), forming a polaron with renormalised energy and mass~\cite{Feynman1955}. Polarons give insight into the transport properties in semiconductors~\cite{PolaronsandExcitonsBook}, electrical conduction in polymer chains~\cite{mahani2017}, spin transport in organic materials~\cite{Gershenson78,Watanabe14} and superconductivity~\cite{Lee06}. Moreover, they can be used as a "probe" for highly correlated quantum many-body environments for instance, $^{3}$He impurities immersed into $^{4}$He superfluids~\cite{LandauFermiLiquidTheory}.\\

Ultracold quantum gases offer a pristine platform where the interaction between atoms can be tuned at will~\cite{Bloch08,ChingCheng2010}, allowing for the possibility to create polarons by immersing atomic impurities in quantum gases. Here, we do not rely on a lattice, but the mechanism is analogous. The impurity gets dressed by the low-energy excitations of the quantum gas instead. Polarons can be created in a Bose-Einstein condensate (Bose polarons) ~\cite{Jorgensen16,hU16,Yan190,Camargo18,Ardila18} or in a degenerate Fermi gas (Fermi polarons)~\cite{Schirotzek09,Nga12,Koschorreck12,Kohstall12,Cetina16,FScazza17,Sidler17}. Albeit research on transport properties using polarons in quantum gases is still premature, these quasi-particles can be used for probing quantum~\cite{Christensen15} and thermal effects~\cite{Mehboudi2019} as well as topological invariants~\cite{GrusdtTop2,Camacho2019,GrusdtTop} in the host bath. However, the precise nature of the dopant is fundamental as the many-body environment can be probed using impurities, rather than polarons~\cite{Bouton2020}, and hence  it is relevant to understand the time scales of polaron formation. Contrary to the condensed-matter scenario, the shortest response to collective excitations in ultracold quantum gases are of the order of microseconds, enabling current state-of-the-art Ramsey techniques to measure these time scales with high precision. Thus, the non equilibrium dynamics of a polaron in quantum gases has attracted a lot of attention both in theory and experiments. In the former case, the dynamics have been investigated using field-theory, renormalization-group, time-dependent variational and perturbative approaches~\cite{Grusdt18r,Volosniev15,Shchadilova16,Lampo2018,Lausch18,Knakkergaard2018,Liu19,Drescher18,Mista1,Mista2}, whereas recent experiments shed light on quasiparticle formation and the non-equilibrium dynamics of impurities, which are quantified by using Ramsey interferometry protocols~\cite{Cetina16sc,Cetina16,Magnus2020}.

In this work, we use a time-dependent variational method~\cite{LLP53,Shchadilova16} to study the real-time evolution of an impurity embedded in a Bose-Einstein condensate. The non equilibrium dynamics is tracked down by measuring the time-dependent contrast, namely, the probability to find the polaron at a certain time with respect to its initial non-interacting state, $\left|\psi(0)\right\rangle$. Formally, the contrast  is defined as $S(t)=\left\langle \psi(0)\right|\exp[-i\hat{H}t]\left|\psi(0)\right\rangle $. Analog quantities are defined  in the context of dynamical quantum phase transitions or quantum chaos, for instance, the Loschmidt echo~\cite{Heyl13,Serbyn17}. Within the variational approach, the contrast is derived for arbitrary times and coupling strengths, which means that one can go beyond the truncated Bogoliubov-Fr\"ohlich Hamiltonian that is accurate only in the weakly interacting regime. 

The article is organized as follows. In Sec. II, we describe the model and we introduce the full Hamiltonian of the system. In Sec. III, we review the time-dependent variational approach to compute the Euler-Lagrange equations of motion (EOM). In Sec .IV,  the EOM can be solved exactly in the weakly interacting regime and a proper benchmark is derived with respect to the perturbative approach~\cite{Knakkergaard2018}. In order to do comparisons with experiments, additional decoherence effects must be included in the theory; in Sec. V, we discuss trap dephasing, and in Sec. VI, decoherence by magnetic-field fluctuations and losses. In Sec. VII, we discuss some results within the experimentally available coupling strengths, both in the weakly and strongly interacting regimes. Comparisons yield a good agreement between the theory and the experimental measurements~\cite{Magnus2020}. Finally, conclusions are drawn in Sec. VIII.

\section{II. Model and Hamiltonian}

\noindent  We consider an ultradilute gas of impurities of mass $m_{I}$ immersed in a Bose-Einstein condensate (BEC) of mass $m_{B}$ and density $n$. The Hamiltonian of the system reads
\begin{eqnarray}
\mathcal{H}&=&\sum_{\mathbf{p}}\Omega_{\mathbf{p}}\hat{c}_{\mathbf{p}}^{\dagger}\hat{c}_{\mathbf{p}}+\sum_{\mathbf{k}}\epsilon_{\mathbf{k}}\hat{a}_{\mathbf{k}}^{\dagger}\hat{a}_{\mathbf{k}} 
\nonumber\\
&+&\frac{1}{2V}\sum_{\mathbf{k,k',q}}V_B(\mathbf{q})\hat{a}_{\mathbf{k+q}}^{\dagger}\hat{a}_{\mathbf{k'-q}}^{\dagger}\hat{a}_{\mathbf{k'}}\hat{a}_{\mathbf{k}}
\nonumber\\
&+&\frac{1}{V}\sum_{\mathbf{k,k',q}}V_I(\mathbf{q})\hat{a}_{\mathbf{k+q}}^{\dagger}\hat{c}_{\mathbf{k'-q}}^{\dagger}\hat{c}_{\mathbf{k'}}\hat{a}_{\mathbf{k}},
\label{eq:HFull}
\end{eqnarray}
where the operator $\hat{c}_{\mathbf{p}}$ ($\hat{c}_{\mathbf{p}}^{\dagger}$) annihilates (creates) an impurity of momentum $\mathbf{p}$ and energy $\Omega_{\mathbf{p}}=\mathbf{p}^{2}/2m_{I}$, whereas $\hat{a}_{\mathbf{k}}$ ($\hat{a}_{\mathbf{k}}^{\dagger}$) annihilates (creates) a boson with momentum $\hbar\mathbf{k}$ and energy $\epsilon_{\mathbf{k}}=\hbar^{2}\mathbf{k^{2}/}2m_{B}$. Moreover, $V_B(\mathbf{q})=T_{B}$ and $V_I(\mathbf{q})=T_{\nu}$ are the Fourier transform of the short-range  boson-boson and impurity-boson potentials, respectively. In addition, we define $T_{\nu}=2\pi\hbar^{2}a/m_{red}$ as the zero-energy impurity-bath coupling constant depending on the tunable \textit{s-}wave scattering length $a$ and  $m_{red}=m_{B}^{-1}+m_{I}^{-1}$ the reduced mass. We perform the Bogoliubov transformation on the bosonic operators, i.e., $\hat{a}_{\mathbf{k}}=u_{k}\hat{b}_{\mathbf{k}}-v_{k}^{*}\hat{b}_{-\mathbf{k}}^{\dagger}$, with amplitudes $u_{k}=1+v_{k}^{2}=\left(\epsilon_{\mathbf{k}}+T_{B}n+\omega_{k}\right)/2\omega_{k}$ and $v_{k}=-T_{B}n/2\omega_{k}$, where  the bosonic coupling strength is $T_{B}=4\pi\hbar^{2}a_B/m_{B}$ and $a_B$ is the \textit{s-}wave boson-boson scattering length. Moreover, $\omega_{\mathbf{k}}=\frac{\hbar^{2}k}{2m_{B}\xi}\sqrt{(k\xi)^{2}+2}$ is the dispersion relation of the bosonic bath written in terms of the healing length, $\xi=1/\sqrt{8\pi na_{B}}$. Thus, the transformed Hamiltonian in the single-impurity limit (characterized by its position operator $\hat{\mathbf{R}}$ ) yields

\begin{eqnarray}
\mathcal{H}&=&\mathcal{H}_{F}+\frac{T_{\nu}}{V}\sum_{\mathbf{k,q}}V_{\mathbf{k,q}}^{(+)}e^{i\left(\mathbf{k-q}\right)\hat{\cdot\mathbf{R}}}\hat{b}_{\mathbf{k}}^{\dagger}\hat{b}_{\mathbf{q}}
\nonumber\\
&+&\frac{T_{\nu}}{V}\sum_{\mathbf{k,q}}V_{\mathbf{k,q}}^{(-)}e^{i\left(\mathbf{k+q}\right)\hat{\cdot\mathbf{R}}}\left(\hat{b}_{\mathbf{k}}^{\dagger}\hat{b}_{\mathbf{q}}^{\dagger}+\hat{b}_{\mathbf{-k}}\hat{b}_{\mathbf{-q}}\right).
\label{eq:BFH}
\end{eqnarray}
 In order to capture all the orders of multiscattering process in the problem, quadratic terms in the Bogoliubov operators must  also be included. Moreover, $V_{\mathbf{k,q}}^{(\pm)}=\frac{1}{2}\left[W_{\mathbf{k}}W_{\mathbf{q}}\pm\left(W_{\mathbf{k}}W_{\mathbf{q}}\right)^{-1}\right]$ with the coupling function  $W_{\mathbf{k}}=\left[\frac{(k\xi)^{2}}{(k\xi)^{2}+2}\right]^{1/4}$ and  $\mathcal{H}_{F}$ reads
\begin{eqnarray}
\mathcal{H}_{F}&=&\frac{\hat{\mathbf{p}}^{2}}{2m_{I}}+\sum_{\mathbf{k}}\hbar\omega_{\mathbf{k}}\hat{b}_{\mathbf{k}}^{\dagger}\hat{b}_{\mathbf{k}}+nT_{\nu}
\nonumber\\
&+&
\frac{T_{\nu}\sqrt{n}}{\sqrt{V}}\sum_{\mathbf{k}}W_{\mathbf{k}}e^{i\mathbf{k}\hat{\cdot\mathbf{R}}}\left(\hat{b}_{\mathbf{k}}^{\dagger}+\hat{b}_{\mathbf{-k}}\right).
\label{eq:Frohlich}
\end{eqnarray}

\noindent The truncated Hamiltonian given by  Eq.~(\ref{eq:Frohlich}) is known as the Bogoliubov-Fr\"ohlich Hamiltonian which provides the ground-state properties of a single impurity weakly interacting with the low-energy excitations of the quantum gas. This process is represented by the scattering of an impurity with a single excitation of the condensate. This truncated Hamiltonian  allows  one to study the impurity dynamics using a  "system plus reservoir " description in terms of a master equation approach (MEA) (see Appendix A). However, it is inaccurate for  strong coupling, which is characterized by a high-energy multiscattering process featuring a transition from low-energy excitations to bare bosons and enabling thus the formation of many-body bound states once the resonance is crossed~\cite{Ardila18,Shchadilova16}. Hence, beyond-Fr\"ohlich terms in Eq.~(\ref{eq:BFH}) are needed for arbitrary coupling strength in order to  properly describe the physics. \\

\textit{Lee-Low-pines (LLP) transformation.} To simplify the problem further, one can translate the impurity-bath system into a new set of coordinates by evoking the celebrated LLP canonical transformation $\mathcal{H}=\hat{U}^{\dagger}\mathcal{H}\hat{U}$, where $\ensuremath{\hat{U}=\exp\left[i\hat{\mathbf{R}}\cdot\sum_{\mathbf{k}}\hbar\mathbf{k}\hat{b}_{\mathbf{k}}^{\dagger}\hat{b}_{\mathbf{k}}\right]}$ ~\cite{LLP53,Shchadilova16,PolaronsandBipolarons}. The transformed Hamiltonian   depends on the total momentum of the system, $U^{\dagger}\mathbf{\hat{P}}U=\mathbf{\hat{P}}-\sum_{\mathbf{k}}\hbar\mathbf{k}\hat{b}_{\mathbf{k}}^{\dagger}\hat{b}_{\mathbf{k}}$. The Hamiltonian given by Eq.~(\ref{eq:BFH})  in the new reference frame reads

\begin{eqnarray}
\mathcal{H}&=&T_{\nu}n+\frac{1}{2m_{I}}\left(\mathbf{P}-\hbar\sum_{\mathbf{k}}\mathbf{k}\hat{b}_{\mathbf{k}}^{\dagger}\hat{b}_{\mathbf{k}}\right)^{2}+\sum_{\mathbf{k}}\omega_{\mathbf{k}}\hat{b}_{\mathbf{k}}^{\dagger}\hat{b}_{\mathbf{k}}
\nonumber\\
&+&T_{\nu}\frac{\sqrt{n}}{\sqrt{V}}\sum_{\mathbf{k}}W_{\mathbf{k}}\left(\hat{b}_{\mathbf{k}}^{\dagger}+\hat{b}_{-\mathbf{k}}\right)+\frac{T_{\nu}}{V}\sum_{\mathbf{k,q}}V_{\mathbf{k,q}}^{(+)}\hat{b}_{\mathbf{k}}^{\dagger}\hat{b}_{\mathbf{q}}
\nonumber\\
&+&\frac{T_{\nu}}{V}\sum_{\mathbf{k,q}}V_{\mathbf{k,q}}^{(-)}\left(\hat{b}_{\mathbf{k}}^{\dagger}\hat{b}_{\mathbf{q}}^{\dagger}+\hat{b}_{\mathbf{-k}}\hat{b}_{\mathbf{-q}}\right).
\label{eq:LLPH}
\end{eqnarray}

 In the new impurity frame of reference, the total momentum commutes with the transformed Hamiltonian and is a conserved quantity. Hence the bare impurity degrees of freedom are fully eliminated and the expectation value of the impurity momentum can be set to zero, which is consistent with the polaron ground state of the system. One more detail that should be taken with care in the Hamiltonian given by  Eq.~(\ref{eq:LLPH}) is the renormalization of the impurity-boson coupling strength as the contact interaction potential is modeled by a nonphysical $\delta$ potential in real space. From the Lippmann-Schwinger equation  the zero-energy coupling constant and the \textit{s-}wave scattering length are related via 

\begin{eqnarray}
T_{\nu}^{-1}=\frac{m_{red}}{2\pi\hbar^{2}a}-\frac{1}{V}\sum_{\mathbf{k}}^{\Lambda\rightarrow\infty}\frac{2m_{red}}{\hbar^{2}\mathbf{k}^{2}},
\label{eq:Couplingconstant}
\end{eqnarray}

\section{III. VARIATIONAL ANSATZ}
\noindent In this section we employ a time-dependent variational ansatz to study polaron dynamics as previously used in Refs. ~\cite{Shchadilova16,LLP53,Drescher18}. On one hand, we benchmark the results obtained within the variational ansatz against a MEA. On the other hand the variational formalism enables us to access  the strongly interacting regime ruled by the full Hamiltonian given by  Eq.~(\ref{eq:LLPH}). The ansatz is based on a coherent state product,

\begin{eqnarray}
\left|\psi_c(t)\right\rangle =\exp[-i\phi(t)]\hat{O}(t)\left|0\right\rangle ,
\label{eq:ansatzs}
\end{eqnarray}

\noindent where $\hat{O}(t)=\exp\left(\frac{1}{\sqrt{V}}\sum_{\mathbf{k}}\beta_{\mathbf{k}}(t)\hat{b}_{\mathbf{k}}^{\dagger}-h.c\right)$ and $\left|0\right\rangle =\left|0_{\mathbf{k}},\mathbf{p}\right\rangle$  is the total state at $t=0$ before  the impurity-boson interaction is quenched, i.e., an impurity with momentum $\mathbf{p}$ and the vacuum of phonons. Here, $\gamma_\mathbf{k}(t)=\left\{ \beta_{\mathbf{k}}(t),\phi(t)\right\}$ plays the role of time-dependent variational parameters. The displacement operator $\hat{O}$ obeys  $\hat{O}^{\dagger}(t)\hat{b}_{\mathbf{k}}\hat{O}(t)=\hat{b}_{\mathbf{k}}+\beta_{\mathbf{k}}$. These properties are heavily used to compute  the equations of motion and  expectation values. Note that, the ansatz Eq.~(\ref{eq:ansatzs}) is exact in the limit of an immobile impurity and non-interacting bosons. Following~\cite{Shchadilova16}, the Euler Lagrange equations, $\frac{d}{dt}\frac{\partial\mathcal{L}}{\partial\dot{\gamma}}-\frac{\partial\mathcal{L}}{\partial\gamma}=0$, with the Lagrangian density, $\mathcal{L}=\left\langle \psi_c(t)\left|i\hbar\partial_{t}-\mathcal{H}\right|\psi_c(t)\right\rangle$ and for the manifold of variational parameters $\gamma_{\mathbf{k}}(t)$ read

\begin{eqnarray}
i\hbar\dot{\beta}_{\mathbf{k}}&=&\frac{T_{\nu}W_{\mathbf{k}}\sqrt{n}}{\sqrt{V}}+\Gamma(\mathbf{k})\beta_{\mathbf{k}}+\frac{T_{\nu}}{2V}\left[W_{\mathbf{k}}\sum_{\mathbf{k}'}W_{\mathbf{k'}}\left(\beta_{\mathbf{k}'}+\right.\right.
\nonumber\\
&+&\left.\left.\beta_{\mathbf{k}'}^{*}\right)+\frac{1}{W_{\mathbf{k}}}\sum_{\mathbf{k}'}\frac{1}{W_{\mathbf{k'}}}\left(\beta_{\mathbf{k}'}-\beta_{\mathbf{k}'}^{*}\right)\right]
\label{eq:betaEq}
\end{eqnarray}
and 
\begin{eqnarray}
\hbar\dot{\phi}=T_{\nu}n+\frac{T_{\nu}\sqrt{n}}{2\sqrt{V}}\sum_{\mathbf{k}}W_{\mathbf{k}}\left(\beta_{\mathbf{k}}+\beta_{\mathbf{k}}^{*}\right)-\frac{\mathbf{P}_{B}^{2}}{2m_{I}}.
\label{eq:gammaEq}
\end{eqnarray}

\noindent Here $\Gamma(\mathbf{k})=\omega_{\mathbf{k}}+\Omega_\mathbf{k}+\hbar \mathbf{k}\cdot \mathbf{P}_{B}/m_{I}$ and $\mathbf{P}_{B}=\sum_{\mathbf{k}}\mathbf{k}\beta_{\mathbf{k}}^{*}\beta_{\mathbf{k}}$. As mentioned previously and without loss of generality, we restrict  ourselves to the case of slow motion impurities, i.e., $\mathbf{P}=0$. Moreover, note that $\mathbf{P_{B}}$ is always zero for all times due to spherical symmetry. Solving the coupled differential equations will allow us to compute quantities of interest such as  the contrast,

\begin{equation}
    S^{0}(t)=\left\langle \psi_c(0)|\psi_c(t)\right\rangle, 
    \label{eq:RamseyGeneral}
\end{equation}

\noindent where $\left|\psi_c(t)\right\rangle$ is given by Eq.~(\ref{eq:ansatzs}). Carrying out the calculation and after some algebra, one ends up with a simple expression ,
\begin{eqnarray}
S^{0}(t)=\exp\left[-i\phi(t)-\frac{1}{2}\sum_{\mathbf{k}}\left|\beta_{\mathbf{k}}(t)\right|^{2}\right].
\label{eq:Ramsey}
\end{eqnarray}

In the stationary case, i.e., $\dot\gamma_{\mathbf{k}}(t)=0$,  Eq.~(\ref{eq:betaEq}) can be split into a real and an imaginary part and hence one obtains

\begin{eqnarray}
\left(\begin{array}{c}
\mathrm{Re}[\beta_{\mathbf{k}}]\\
\mathrm{Im}[\beta_{\mathbf{k}}]
\end{array}\right)=\left(\begin{array}{c}
-\frac{W_{\mathbf{k}}\sqrt{n}}{\Omega_\mathbf{k}}\left(T_{\nu}^{-1}+\sum_{\mathbf{k'}}\frac{\left|W_{\mathbf{k'}}\right|^{2}}{\Omega_\mathbf{k'}}\right)^{-1}\\
0
\end{array}\right).
\label{eq:ReIm}
\end{eqnarray}

Thus, the polaron energy in the stationary case can be computed by  $E_{pol}=\left\langle \psi_{st}\left|\mathcal{\widehat{H}}\right|\psi_{st}\right\rangle$, yielding

\begin{eqnarray}
E_{pol}=\frac{8\pi(na_{B}^{3})^{1/3}}{\left(6\pi^{2}\right)^{2/3}}\left[\frac{a_{B}}{a}-\frac{a_{B}}{a_{0}}\right]^{-1}\frac{\hbar^{2}k_{n}^{2}}{2m_{B}}.
\label{eq:E0var}
\end{eqnarray}

\begin{figure}
\begin{center}
\includegraphics[width=7.5cm]{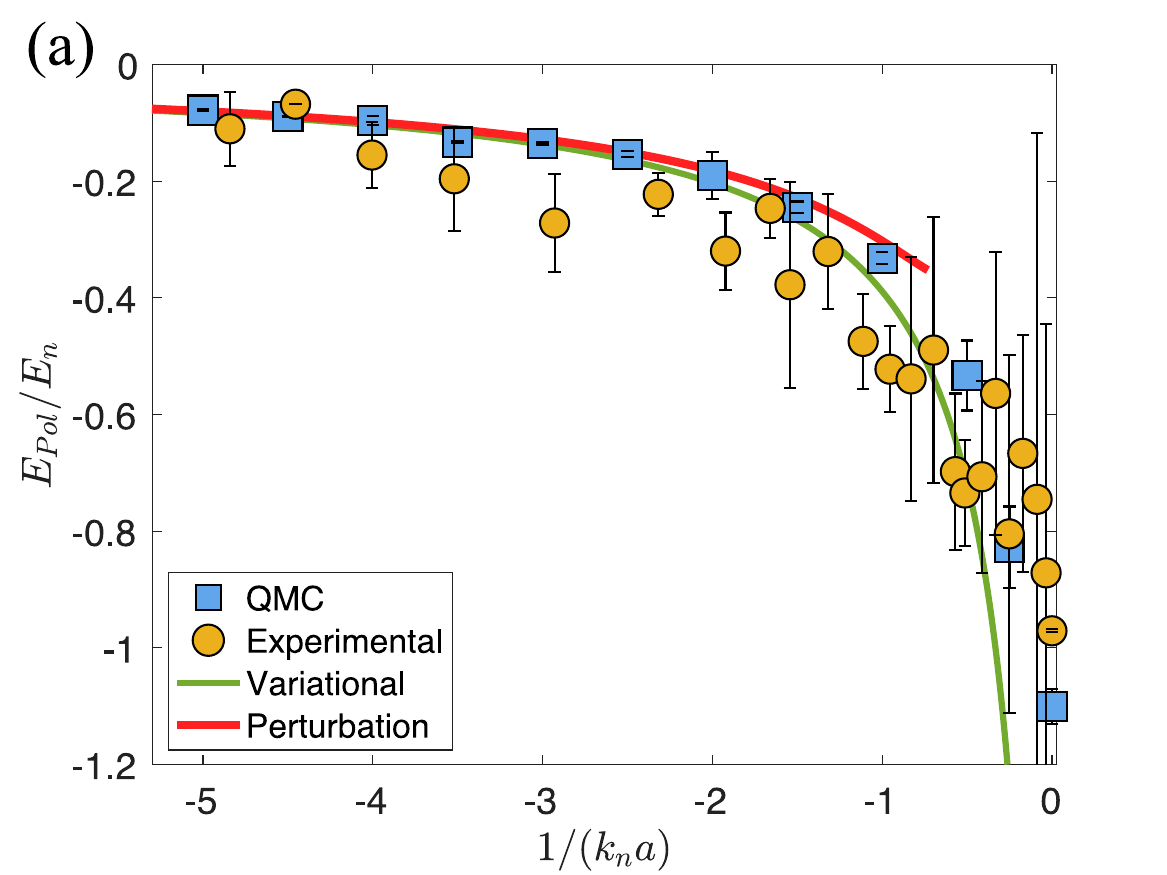}
\includegraphics[width=7.2cm]{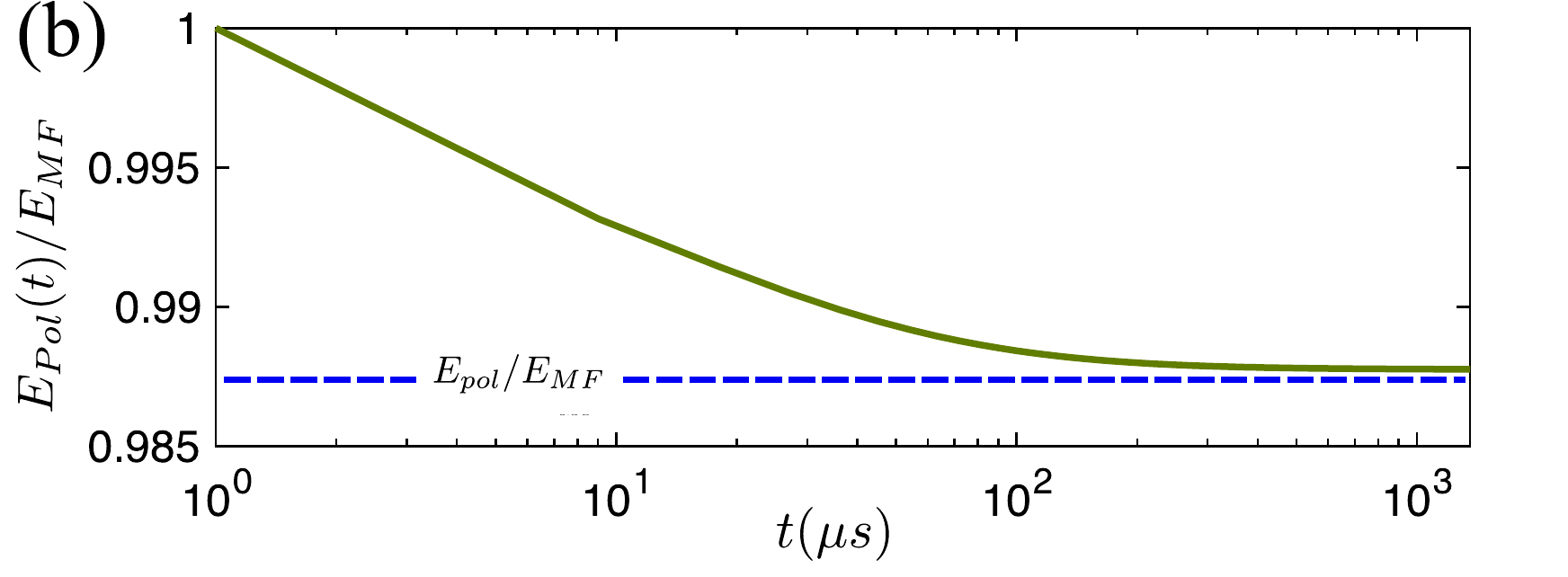}
\caption{(a) Polaron energy at equilibrium as a function of the coupling strength $1/k_{n}a$ using quantum Monte Carlo (blue square) and experimental  (orange circle) data  from Ref. ~\cite{Ardila18}, second-order perturbation theory (red curve) given by  Eq.~(\ref{EFrohlich}) in the Appendix, and time-dependent variational ansatz (green curve) given by Eq.~(\ref{eq:E0var}). (b) Example of instantaneous polaron energy $E_{Pol}(t)=\left\langle \psi_{c}(t)\left|\mathcal{H}\right|\psi_{c}(t)\right\rangle$  as a function of time for a coupling strength $1/k_na=-5$. The polaron is expected to reach equilibrium with an energy $E_{Pol}$ [see Eq.~(\ref{eq:E0var})].}
\label{fig1}
\end{center}
\end{figure}

Alternatively, one can check that for long times , $\left|S(t\rightarrow\infty)\right|\rightarrow Z=\exp\left(-\frac{1}{2V}\sum_{\mathbf{k}}\left|\beta_{\mathbf{k}}\right|^{2}\right)$, which coincides  with the amplitude of the contrast for the homogeneous case and, in particular, at weak coupling, one recovers the quasiparticle residue predicted by perturbative approaches~\cite{Christensen15}. As aforementioned, the expression for the energy given by  Eq.~(\ref{eq:E0var}) is, in principle, valid for arbitrary coupling strength; in fact,  one recovers  the perturbative result [see Eq.~(\ref{EFrohlich}) in Appendix A] for $a/\xi\ll1$  . In order to check the quality of the ansatz against other theories and  the experiment, in Fig.~\ref{fig1} we compare the polaron energy  using  recent experimental data and quantum Monte Carlo methods ~\cite{Ardila18} against the variational result in Eq.~(\ref{eq:E0var}). The agreement is quite good within statistical error, even for large values of $1/k_{n}a$. Nonetheless,  near-unitarity differences might be significant. In fact, the variational ansatz relies on the Bogoliubov approximation which is well known to be less accurate close to the resonance; however, for time scales far enough from the  polaron stationary state the variational method still provides a quantitatively good description of the impurity decoherence.



\section{IV. Decoherence-weak coupling}
The coupled system of Eqs.~(\ref{eq:betaEq}) and~(\ref{eq:gammaEq}) is computed using the full Hamiltonian in Eq.~(\ref{eq:LLPH}) and should be solved numerically. However, for weak coupling, one can use the truncated Fr\"ohlich Hamiltonian  given by  Eqs.~(\ref{eq:Frohlich}), and the EOM are reduced to
\begin{eqnarray}
i\hbar\dot{\beta_{\mathbf{k}}}&=&\frac{T_{\nu}}{\sqrt{V}}\sqrt{n}W_{\mathbf{k}}+\Gamma(\mathbf{k})\beta_{\mathbf{k}}
\nonumber\\
\hbar\dot{\phi}&=&T_{\nu}n+\frac{T_{\nu}\sqrt{n}}{2\sqrt{V}}\sum_{\mathbf{k}}W_{\mathbf{k}}\left(\beta_{\mathbf{k}}+\beta_{\mathbf{k}}^{*}\right).
\label{eq:EL_W}
\end{eqnarray}

The previous equations can be solved exactly. Notice that, in the mean-field case, the solution for the contrast is trivial since $\beta_{\mathbf{k}}(t)=0$ and $\phi(t)=\frac{T_{\nu}n}{\hbar}t$, and hence the contrast reads

\begin{equation}
    S^{h}(t)=\exp\left[-iT_{\nu}nt/\hbar\right].
    \label{eq:meanfield_s}
\end{equation}

In order to recover beyond mean-field terms, the strategy is to solve the first equation in Eq.~(\ref{eq:EL_W}) for $\beta_{\mathbf{k}}$. It yields $\beta_{\mathbf{k}}(t)=\frac{T_{\nu}}{\sqrt{V}}\sqrt{n}\frac{W_{k}}{\Gamma(\mathbf{k})}\left[\exp\left(-i\Gamma(\mathbf{k})t/\hbar\right)-1\right]$ for the initial condition $\beta_{\mathbf{k}}(0)=0$. Afterward this solution is plugged into the $\phi-$equation, thus obtaining   $\phi=\frac{1}{\hbar}\int_{0}^{t}ds\left[T_{\nu}n+\frac{T_{\nu}\sqrt{n}}{2\sqrt{V}}\sum_{\mathbf{k}}W_{\mathbf{k}}\left(\beta_{\mathbf{k}}(s)+\beta_{\mathbf{k}}^{*}(s)\right)\right]$. In addition, one should take into account the proper renormalization of the impurity-boson coupling strength, $T_{\nu}$, by means of Eq.~(\ref{eq:Couplingconstant}). One can easily arrive at a closed form for the variational parameters  $\beta_{\mathbf{k}}(t)$ and $\phi(t)$. Plugging  the previous solutions into the definition of the contrast given by  Eq.~(\ref{eq:Ramsey}), one finally gets $S^{h}(t)=\exp\left[-iE_{pol}^{F}t+i\frac{1}{8}\frac{a_{B}}{a_{0}}\left(\frac{a}{a_{B}}\right)^{2}F(t/t_{n})\right]$,  which coincides with the  expression found by the MEA in Appendix  A, Eq.~(\ref{eq:contrastweak}). The variational approach surprisingly captures  the quantitative behavior of the polaron dynamics out of equilibrium. This conclusion yields interesting consequences. In Ref. ~\cite{Knakkergaard2018}, the master equation for the impurity problem was derived upon a strong assumption: the Markov-Born approximation, which was genuinely justified by the fact that the impurity weakly perturbs the host medium contrary to what is predicted in ~\cite{Lampo2018}  where the memory effects are relevant. Nonetheless, in the current experimental conditions these effects seems to play a small role in the polaron dynamics.
	
\section{V. Trap dephasing}


\begin{figure*}
\begin{centering}
\includegraphics[width=7.5cm]{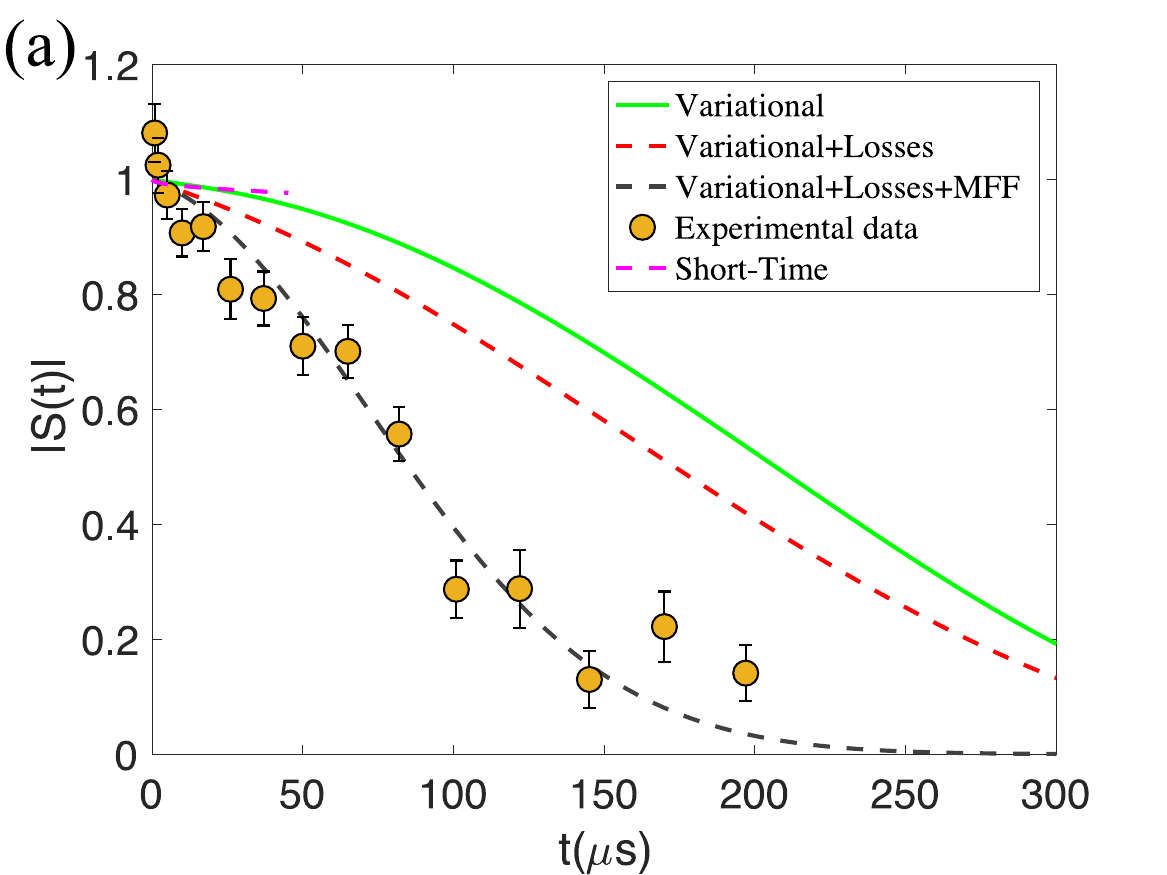}
\includegraphics[width=7.5cm]{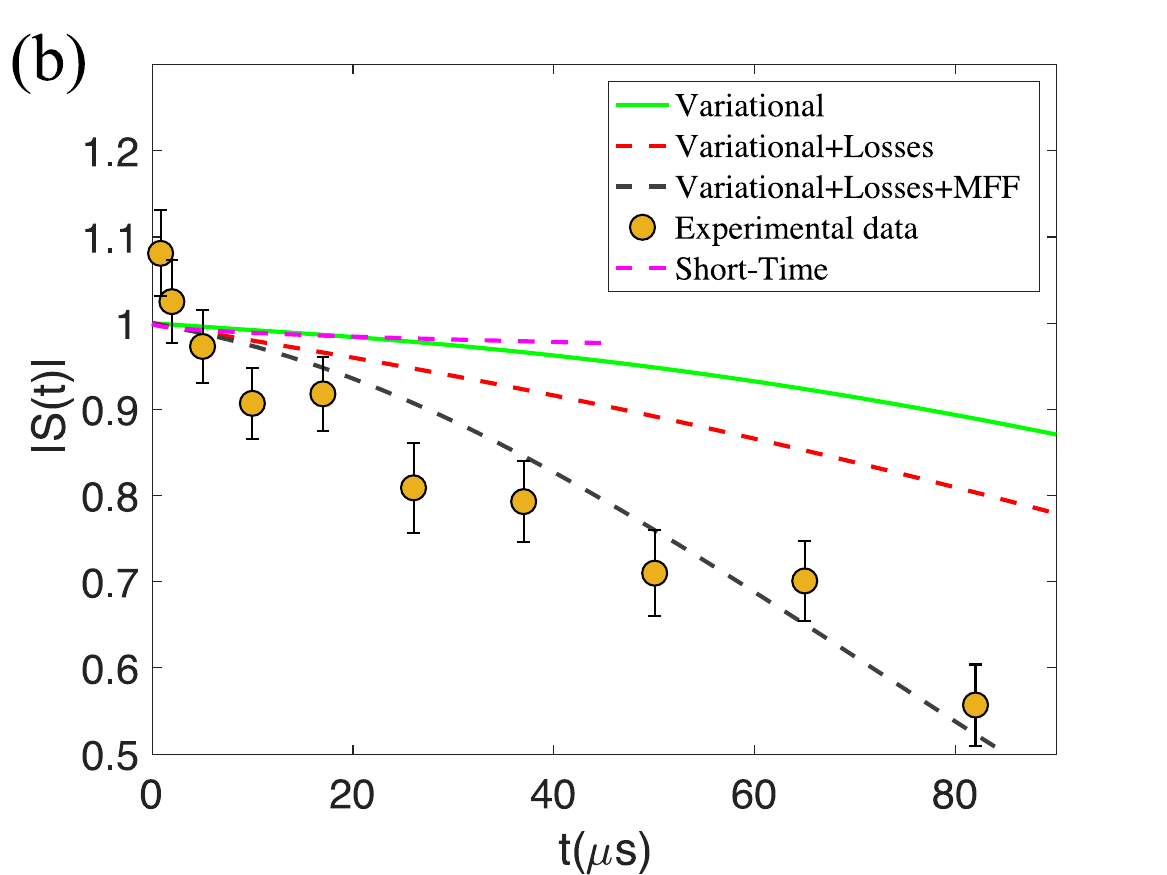}

\caption{(a) Contrast amplitude for $1/k_{n}a=-5.0$. The experimental points from \cite{Magnus2020} are represented by orange circles. The variational result is depicted by the green line and overlaps with the  second-order perturbative result. The purple dashed line depicts  the short-time behavior characterized by the time scale  $\sqrt{t_n}$  (defined in the  text). On top of  trap dephasing decoherence, the variational approach also includes decoherence by losses represented by the dashed red  line and also includes the  magnetic-field fluctuations (MFFs) depicted by the dashed black line (b) zoom-in of (a) in order to  better distinguish  all theories for short and intermediate times. The MFF fitting experimental parameter here is $\Delta=1800$Hz.}
\label{fig2}
\end{centering}
\end{figure*}

If we consider the simplest case at the mean-field level, the amplitude of the contrast given by  Eq.~(\ref{eq:meanfield_s}) is one for all times. In order to do a proper comparison with the experiments, the contrast must be averaged over a non homogeneous density profile, i.e. dephasing by the trap. In the mean-field case and using Eq.~(\ref{eq:meanfield_s}) one finds 

\begin{eqnarray}
S(t)=\frac{1}{N_{B}}\int d^{3}\mathbf{r}n(\mathbf{r})\exp\left[-iT_{\nu}n(\mathbf{r})\frac{t}{\hbar}\right],
\label{eq:meanfieldtotal}
\end{eqnarray}

where  $n(\mathbf{r})=n_{\mathrm{TF}}(\mathbf{r})$  is the condensate density (see Appendix B), whereas the first beyond-mean-field correction reads

\begin{eqnarray}
S(t)=\frac{1}{N_{B}}\int d^{3}\mathbf{r}n(\mathbf{r})S(\mathbf{r},t),
\label{eq:secondordertotal}
\end{eqnarray}
with 
\begin{equation}
   S(\mathbf{r},t)=\exp\left[-\frac{i}{\hbar}E_{pol}^{F}\left[n(\mathbf{r})\right]t+i\frac{1}{8}\frac{a_{B}}{a_{0}}F(t/t_{n(\mathbf{r})})\left(\frac{a}{a_{B}}\right)^{2}\right],
\end{equation}

\noindent containing the inhomogeneity of the condensate in terms of the local time scale $t_{n(\mathbf{r})}=\frac{m_{B}}{8\pi n(\mathbf{r})a_{B}\hbar}$.

\section{VI. Dephasing by losses and  magnetic field fluctuations (MFF)}

 Inelastic losses arising from the three-body recombination process can strongly influence the contrast and give rise to decoherence by losses. The experimental impurity loss rate as a function of the coupling strength has been measured in ~\cite{Magnus2020} using the function $\frac{\hbar\Gamma_{\mathrm{loss}}}{E_{n}}=A\exp(\frac{B}{k_{n}a})+C$, with fitting parameters  $A=0.1788$, $B=1.0130$, and $C=0.0061$. In addition, $\hbar/E_{n}=2m/\left(\hbar^{3/2}6\pi^{2}n\right)^{2/3}$. On the other hand, the Ramsey interferometry protocol is sensible to magnetic -field  variations for each experimental measurement and hence MFF is another important source of decoherence~\cite{Magnus2020}. Therefore, the total contrast needs to also include  these additional inherent effects from the experiment. Thus, the total contrast is modified as

\begin{eqnarray}
S(t)=\frac{1}{N_{B}}S^{\mathrm{Loss}}(t)S^{\mathrm{MFF}}(t)\int d^{3}\mathbf{r}n(\mathbf{r})S(\mathbf{r},t),
\label{eq:ContrastLoss}
\end{eqnarray}

where

\begin{eqnarray}
S^{\mathrm{Loss}}(t)=\exp\left[-\Gamma_{\mathrm{Loss}}t\right]
\label{eq:ContrastLoss1}
\end{eqnarray}

and

\begin{eqnarray}
S^{\mathrm{MFF}}(t)=\frac{\int_{-\infty}^{\infty}d\phi\exp\left[-i\phi-\frac{\phi^{2}}{8\pi^{2}\varDelta^{2}t^{2}}\right]}{\sqrt{8\pi^{3}\varDelta^{2}t^{2}}},
\label{eq:ContrastLoss2}
\end{eqnarray}

are the decoherence associated to losses and MFFs, respectively. Instead, $S(\mathbf{r},t)$ is the contrast including only the trap dephasing. From the experiments, one observes that dephasing due to the trap and by losses are the most important contributions on top of the dephasing by collisions. Additional decoherence effects, such as the one due to magnetic field fluctuations (MFFs) are quite relevant in the weakly interacting regime and become  less important for strong coupling, albeit never negligible.

\section{VII. Results}

\textit{Weak coupling.} The time evolution of the contrast is plotted in Fig.~\ref{fig2}. The circles with error bars depict the recent experimental data~\cite{Magnus2020} for a fixed coupling strength $1/k_{n}a=-5.0$, featuring a typical value in the weakly interacting regime. The computed contrast using the variational ansatz approach including only trap dephasing is represented by the green solid curve. This  result matches with the predicted contrast using a MEA. In addition, the dashed purple line shows the power-law short-time behavior [see, also,  Eq.~(\ref{eq:-2}) in Appendix A] that is valid for $t_{d}<t\ll t_{n}=\hbar/E_{MF}\sim 50\mu$s for this particular coupling strength. Note that  the short-time prediction in Fig.~\ref{fig2} represents the initial high-energy two-body dynamics ~\cite{Knakkergaard2018},  which is independent of the host bath interaction, i.e.,  the impurity does not experience the condensate presence. Therefore, decoherence is originated only by impurity-boson interactions. This regime is preceded by the polaron relaxation and formation stage, where the impurity interacts with the condensate excitations and many-body effects start to take place ($t\approx t_{n}$). Instead, the polaron formation  is expected for times $t\gg t_B \sim 500 \mu$s as revealed by the instantaneous polaron energy defined as $E_{Pol}(t)=\left\langle \psi_{c}(t)\left|\mathcal{H}\right|\psi_{c}(t)\right\rangle=\hbar\dot{\phi}$ and plotted in Fig.~\ref{fig1}(b). A discussion of the degeneracy time scale $t_{d}=ma^{2}/\hbar$ can be found in \cite{Braaten10,Magnus2020}. This time scale is set by the coupling strength and, for this particular case, is of the order of $t_{d}=0.08\mu$s.

For very short times $t_d<t<t_{n}$, the mean-field prediction already provides a good qualitatively agreement with respect to the  experimental contrast; however, the effect of quantum fluctuations (originated from impurity-bath collisions) tends to shift the mean-field contrast amplitude for larger times, following thus the  experimental downward trend as shown in Figs. ~\ref{fig2}(a) and \ref{fig2}(b), but still far from a fully quantitatively agreement with respect to the experimental data. For intermediate and larger times  $t_{n}>100 \mu$s, the contrast decays faster in the experiment with respect to the theoretical prediction which only includes trap dephasing. In fact, additional decoherence effects which are inherent to the experiment are absent in the theory and need to be included, for instance, decoherence by losses and by MFFs ~\cite{Magnus2020}. We have included the losses dephasing  as well as the MFFs decoherence as illustrated in Sec. VI. A quantitative agreement with respect to the experimental data is reached as shown by the black dashed line in Figs. ~\ref{fig2}(a) and ~\ref{fig2}(b). The effect of including the  losses' dephasing is more evident in the strongly interacting regime, in contrast to the dephasing due to MFFs which is less relevant.\\

\begin{figure*}
\begin{centering}
\includegraphics[width=7.5cm]{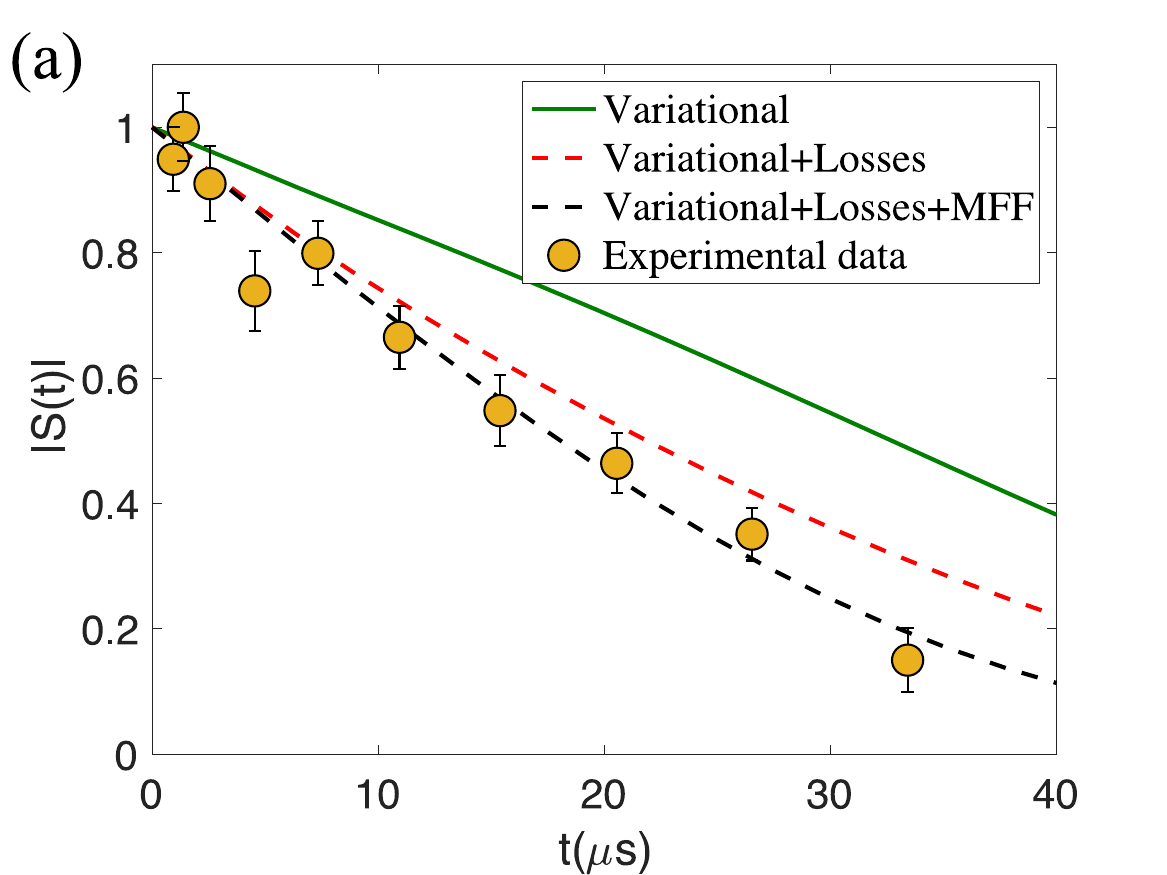}
\includegraphics[width=7.5cm]{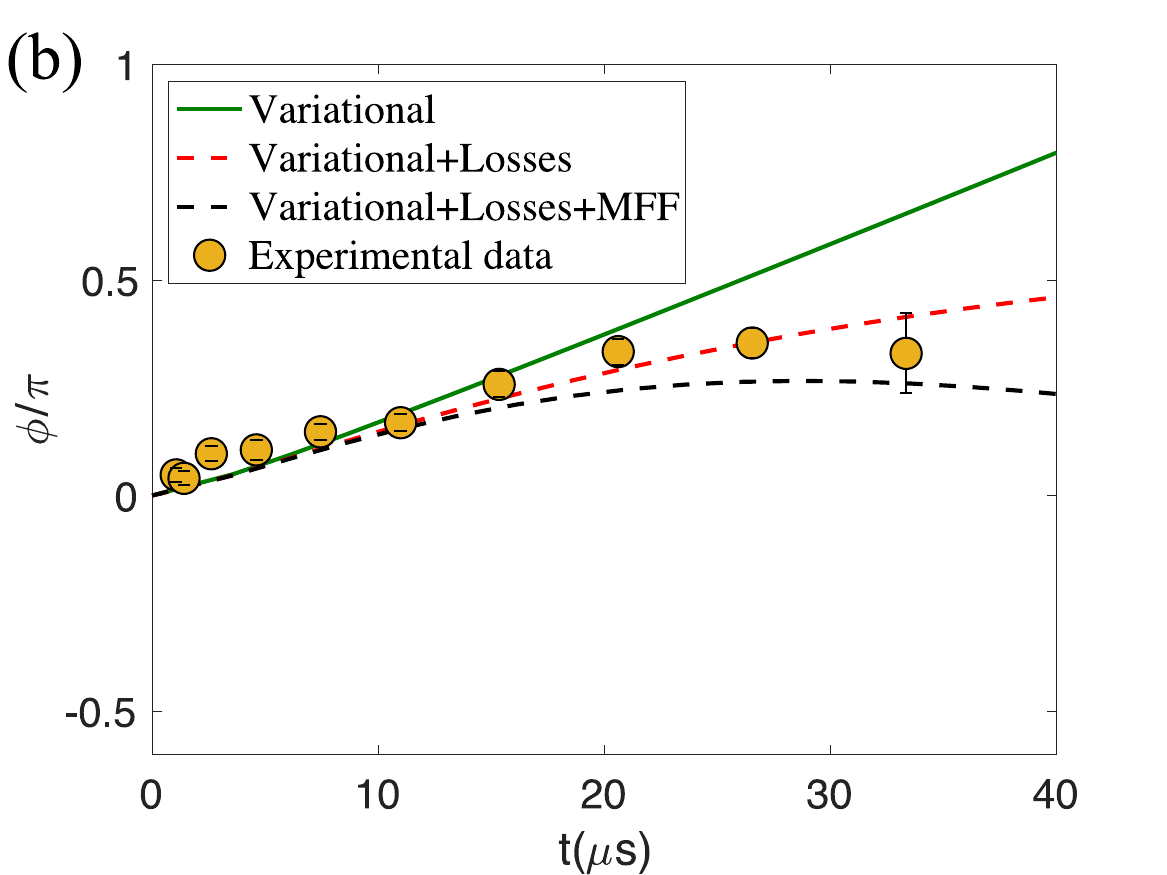}
\includegraphics[width=7.5cm]{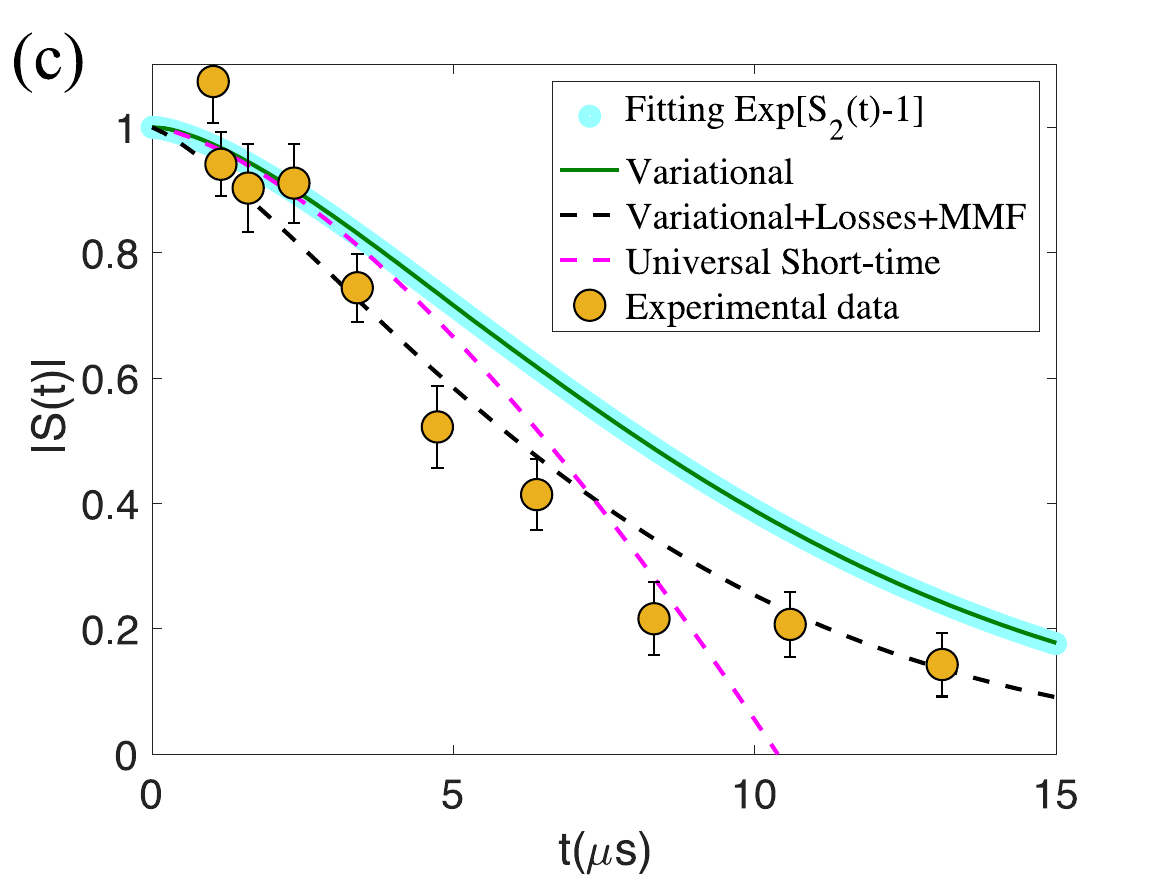}
\includegraphics[width=7.5cm]{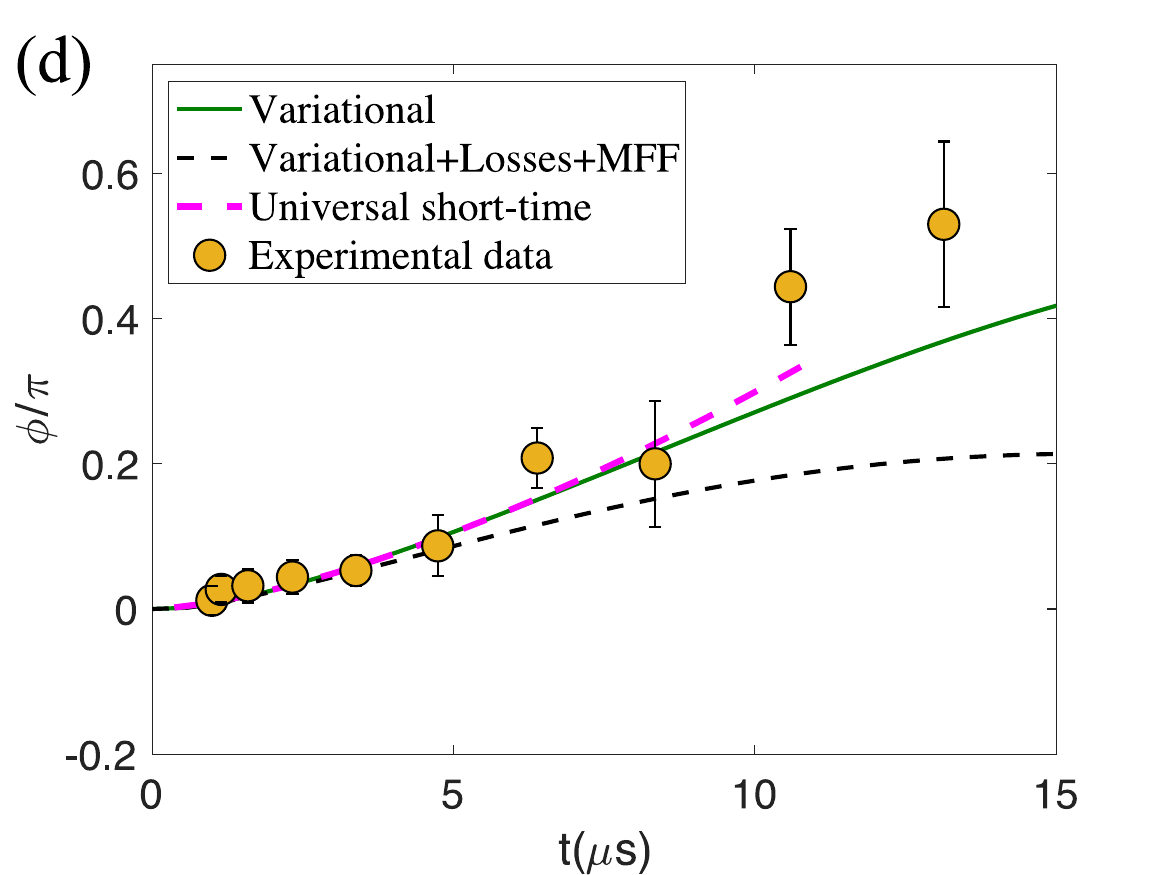}
\caption{(a),(b) Amplitude and phase of $S(t)$  for  $1/(k_{n}a)=-1$. Orange points represent the experimental measurements in~\cite{Magnus2020}. The green line depicts  the variational calculation including only trap dephasing, whereas the dashed lines include the losses dephasing (red line) and, on top, MFFs (black line). (c),(d) Amplitude and phase of $S(t)$ at unitarity. In addition,  the cyan line indicates the fitting proposed in Eq.~(\ref{eq:B}) and the magenta dashed line depicts  the short-time universal behavior. The MFFs fitting parameter here is $\Delta=4600$ Hz.}
\label{fig3}
\end{centering}
\end{figure*}

\textit{Strong coupling.} We investigate the time-dependent contrast in the strongly interacting regime, i.e., $-1<\left(k_{n}a\right)^{-1}<0$. Here we rely solely on numerical solutions of  Eqs.~(\ref{eq:betaEq}) and~(\ref{eq:gammaEq}) with the experimental parameters associated to each $1/k_{n}a$~\cite{Magnus2020}; in particular, for two characteristic coupling strengths, $1/k_{n}a=-1$, and  at unitarity. In the former case, a very good agreement between the variational results including the  losses'  decoherence,  MFFs, and the experimental data is achieved, as shown in Figs.~\ref{fig3}(a) and \ref{fig3}(b).  In this regime, MFF decoherence is small in comparison with the other experimental decoherence effects; however, it provides a quantitative agreement with the experimental contrast, whereas at unitarity it is practically negligible.\\

\begin{figure}
\begin{centering}
\includegraphics[width=10.0cm]{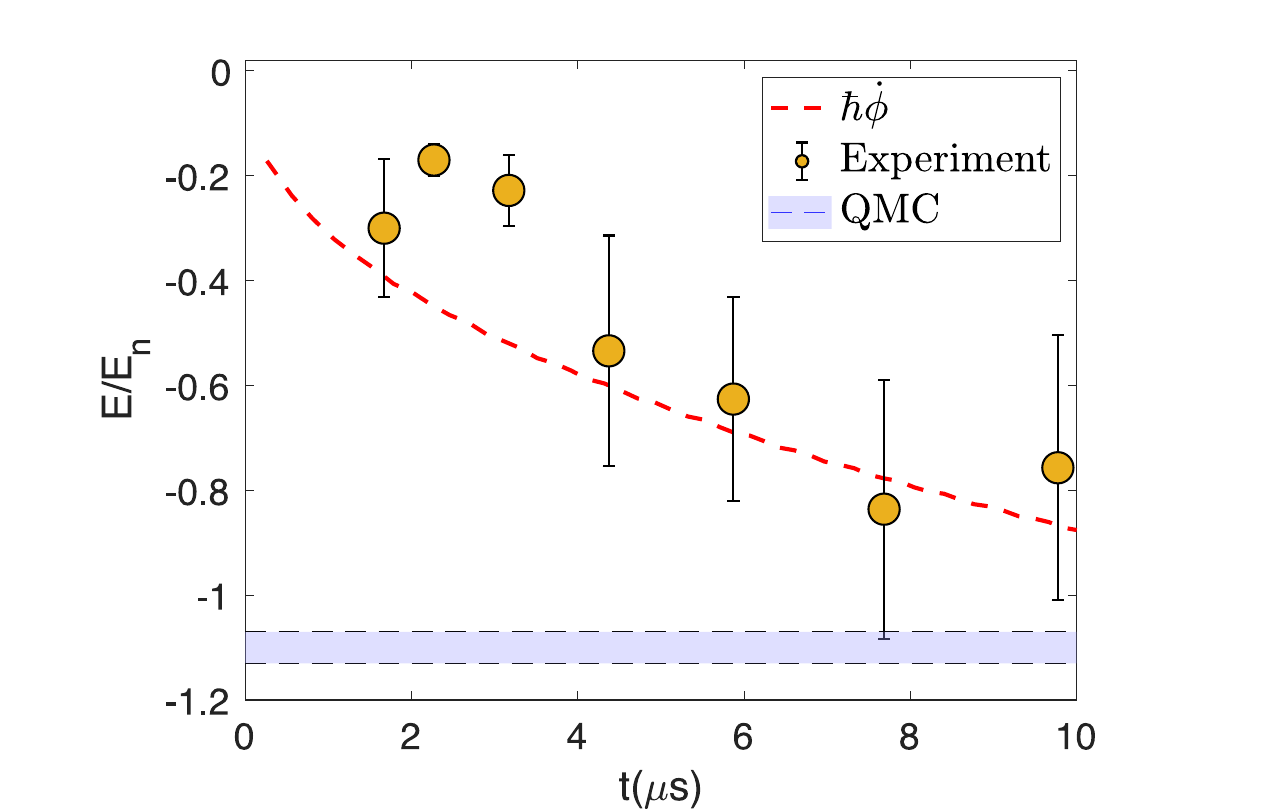}
\caption{Initial departure of the instantaneous polaron energy computed with the variational approach (red dashed line) against the experimental measurements (orange points)~\cite{Magnus2020} at unitarity. The blue region is the quantum Monte Carlo (QMC) prediction for the polaron stationary energy~\cite{Ardila18}.}
\label{fig4}
\end{centering}
\end{figure}

In Figs.~\ref{fig3}(c) and \ref{fig3}(d), we show the results at unitarity. We observe that both the initial relaxation obtained within the variational ansatz (green line) and the short-time universal behavior~\cite{Braaten10}, valid for $t\ll t_{d}$ (magenta dashed line),  agree with each other and it follows that
\begin{eqnarray}
S_{\mathrm{2body}}(t)\approx1-(1-i)\frac{16}{9\pi^{3/2}}\left(\frac{t}{t_{n}}\right)^{3/2}.
\label{eq:universalunit}
\end{eqnarray}
 The time-scale  $t_{n}^{3/2}$ is linked to the universal two-body scattering physics, where the impurity interacts with  highly energetic bosons of the condensate. Note that  the same universal scaling is recovered for impurities immersed in a Fermi sea~\cite{Parish2016} with a characteristic time-scale set by the Fermi time ($\hbar/E_{F}$). In order to bridge the two-body with the many-body physics where the polaron starts its formation, we attempt to find a functional form of the contrast for arbitrary times. Taking advantage of the functional form of Eq.~(\ref{eq:universalunit}), we propose the following expression for arbitrary time:
\begin{eqnarray}
S(t)=\exp\left[S_{\mathrm{2body}}(t)-1\right],
\label{eq:B}
\end{eqnarray}

which recovers exactly Eq.~(\ref{eq:universalunit}) for short times and spans  over all  one-body, two-body,  and  high $n-$ body correlations for larger times, thus signaling the transition from  few-body to the many-body correlated regime (around $t\sim5\mu$s). The agreement between Eq.~(\ref{eq:B}) depicted by the cyan curve in Fig.~\ref{fig3}(c) and the full numerical calculations obtained with the variational ansatz (green curve) is  remarkable ~\cite{SM}. Interestingly, in Fig.~\ref{fig4}, one observes that the polaron is expected to enter the equilibrium regime for times $t\gg 5\mu$s with a polaron energy near to $E_{Pol}\approx-E_{n}$ at equilibrium. In addition, the proposed function is heavily based on the initial form of the ansatz given by  Eq.~(\ref{eq:ansatzs}), which encodes  all correlations of the system. In fact, by expanding out the wave function Eq.~(\ref{eq:ansatzs}), one has 
\begin{eqnarray*}
\left|\psi(t)\right\rangle &=&\exp\left(-i\phi(t)\sum_{\mathbf{k}}\left[\beta_{\mathbf{k}}(t)\hat{b}_{\mathbf{k}}^{\dagger}-\beta_{\mathbf{k}}^{*}(t)\hat{b}_{\mathbf{k}}\right]\right)\left|0\right\rangle 
\nonumber\\
&=&\sum_{n=0}^{\infty}\frac{\left(-i\phi(t)\left[\sum_{\mathbf{k}}\beta_{\mathbf{k}}(t)\hat{b}_{\mathbf{k}}^{\dagger}-\beta_{\mathbf{k}}^{*}(t)\hat{b}_{\mathbf{k}}\right]\right)^{n}}{n!}\left|0\right\rangle ,
\label{eq:D}
\end{eqnarray*}

\noindent and taking into account only the first terms of the series, one obtains

\begin{equation}
\left|\psi(t)\right\rangle \simeq\left(1-i\phi(t)\sum_{\mathbf{k}}\beta_{\mathbf{k}}(t)\hat{b}_{\mathbf{k}}^{\dagger}\right)\left|0\right\rangle,
 \label{chevysum}
\end{equation}

which resembles the Chevy ansatz~\cite{Li14} where only a single excitation on top of the unperturbed condensate is included. Equation~(\ref{chevysum}) is accurate in the weakly interacting regime. Hence, perturbation theory and the variational approach again consistently agree in this regime.\\
Finally, we would like to stress that in the neighborhood of  unitarity and for very long time (essentially the polaron state at equilibrium), the variational approach might provide quantitative results as in this regime the Bogoliubov approximation breaks down. However, for our current work, the time scales are very far from equilibrium at unitarity and hence good qualitative results in the short-time dynamics hold.

\section{VIII. Conclusions}

In this work, we have employed a time-dependent variational approach  to investigate the out-of-equilibrium dynamics of impurities interacting with a Bose-Einstein condensate. The main results are two fold. First, the method is benchmarked against time-dependent perturbative approaches such as MEA, valid in the weakly interacting regime. Second, the method is compared with state-of-the-art experimental data. Besides impurity-bath decoherence, agreement between theory and experiment relies on the inclusion of decoherence effects in the theoretical approaches due to trap inhomogeneity, as well as  losses and fluctuations of the magnetic field inherently incorporated in the experiment. Close to  unitarity, we also bridged, in a consistent and quantitative way, the few-body physics characterized by the universal two-body scattering process to the many-body process where polaron starts its formation. Note that  apart from the initial experimental preparation of the system, the contrast might be altered due to the comparable Efimov states'  energy that is also relevant for polaron physics in bosonic systems~\cite{Levinsen15}. Also, how the polaron formation happens once the resonance is crossed is an open question since a new time scale related with the two-body bound state physics should play an important role.
Real-time formation of quasiparticles emerging from impurity-bath decoherence might be important not only for gases, but also for other exotic many-body environments such as ultra dilute quantum liquids or the recently investigated long-range polarons such as Rydberg, dipolar, and ionic polarons~\cite{Bisset2020,DipolarPolaron18,Kain14,Camargo18,astrakharchik2020ionic}.\\

\textit{Note added--.} Recently, a study of exact quench dynamics of the ideal Bose polaron at zero and nonzero temperatures~\cite{Drescher21}  derived a very similar expression for the contrast in our  Eq.~(\ref{eq:B}).

\section*{Acknowledgments}
We gratefully acknowledge  Kristian K. Nielsen for discussions at an earlier stage of this work. We also thank  Magnus G. Skou, Nils Byg J{\o}rgensen and Jan Artl for insightful discussions and for providing the experimental data. The authors also acknowledge  Luis Santos, Fabian Grusdt, Artem Volosniev, and Giacomo Bighin for reading the manuscript and for the critical feedback. Special acknowledgment is given to Roberta Giusteri for critical comments on the manuscript. This research was funded by the DFG Excellence Cluster QuantumFrontiers.

\section{Appendix A: Master Equation approach (MEA)}

At $t=0$, the total density matrix of the whole system can be factorized as the product of the  density matrix of a small subsystem (impurities) and the host reservoir (bath). In the spirit of the Born approximation, when the interaction system bath is turned on, correlations play an important role giving place to decoherence of the polaron and hence the density matrix of the system displays deviations on the order of the system-reservoir coupling strength~\cite{MasterEquationsFokker-Planck}. This situation can be described accurately within a MEA, provided that interactions (impurity-bath) are weak enough~\cite{Knakkergaard2018}. In fact, Eq.~(\ref{eq:Frohlich}) in the main text follows a system-reservoir kind of Hamiltonian. In order to characterize the coherence of the system, we study the time-dependent contrast (also known as Ramsey overlap), $S(t)=\left\langle 0\left|\hat{c}(t)\rho_{I}(t)\right|0\right\rangle$, where $\rho_{I}(t)$ is the reduced density matrix of the impurity obtained explicitly with the MEA. In Ref.~ \cite{Knakkergaard2018} a detailed  derivation of $\rho_{I}(t)$ is discussed. Moreover, $\hat{c}(t)$ depicts the impurity operators and $\left|0\right\rangle $ is the vacuum of phonons. Thus within the Fr\"ohlich model, the decoherence for a uniform system yields \cite{Knakkergaard2018}

\begin{equation}\tag{A1}
S^{h}(t)=\exp\left[-\frac{i}{\hbar}E_{pol}^{F}t+i\frac{1}{8}\frac{a_B}{a_{0}}F(t/t_n)\left(\frac{a}{a_{B}}\right)^{2}\right],
\label{eq:contrastweak}
\end{equation}

\noindent with $t_{n}=m/8\pi\hbar na_{B}$ and the function $F(t)=t+i\left(it+3\right)\left(\frac{1+i}{2}\right)\sqrt{\pi t}\exp\left(it/2\right)\mathrm{erfc}\left[\left(\frac{1+i}{2}\right)\sqrt{t}\right]$. The upper index $``h"$ indicates the contrast computed for a homogeneous system and

\begin{equation}\tag{A2}
    E_{pol}^{F}=\frac{8\pi}{(6\pi^{2})^{2/3}}(na_{B}^{3})^{1/3}\left[\frac{a}{a_{B}}+\frac{a_{B}}{a_{0}}\left(\frac{a}{a_{B}}\right)^{2}\right]\frac{\hbar^{2}k_{n}^{2}}{2m}
    \label{EFrohlich}
\end{equation}

\noindent is the polaron energy in the Fr\"ohlich regime. Here, $k_{n}=\left(6\pi^{2}n\right)^{1/3}$ and $\frac{a_{B}}{a_{0}}=\frac{32}{3\sqrt{\pi}}\sqrt{na_{B}^{3}}$. In addition, for short time, namely, $t\ll\hbar/E_{\mathrm{MF}}$ (where $E_{\mathrm{MF}}=T_{\nu}n$ is the mean-field polaron energy), the contrast yields a particular form

\begin{equation}\tag{A3}
S^{0}(t)\sim1-(i+1)\sqrt{\frac{t}{t_{\Omega}}}-\frac{i}{\hbar}E_{\mathrm{MF}}t,
\label{eq:-2}
\end{equation}

\noindent with the characteristic time-scale $t_{\Omega}=\frac{m_{B}}{32\pi\hbar n^{2}a^{4}}$, which depends on the shape of the impurity-bath potential. Both theoretical approaches addressed in this work assume a quenched impurity.  Universality of the very short time dynamics and the corresponding time scales are discussed in ~\cite{Magnus2020} and  are not within the scope of the present work.

\section{Appendix B: contrast for an inhomogeneous trap}

The density of the condensate can be written in a very simple form as the interaction energy scales are larger with respect to the kinetic energy one, and hence the density of the condensate within the Thomas-Fermi (TF) approximation can be written as

\begin{eqnarray*}
n_{\mathrm{TF}}(\mathbf{r})=\left\{\begin{array}{c}
\frac{\mu_{\mathrm{TF}}-V(\mathbf{r})}{T_{B}},\quad\mu_{\mathrm{TF}}\ge V(\mathbf{r})\\
\\
0,\quad\mathrm{Otherwise},
\end{array}\right.
\label{eq:density}
\end{eqnarray*}

with $\mu_{\mathrm{TF}}$ the chemical potential in the TF regime. In experiments, the trapping potential is modeled by $V(\mathbf{r})=\frac{1}{2}m_{B}\sum_{i=1}^{3}\omega_{i}^{2}x_{i}^{2}=\frac{1}{2}m_{B}\overline{\omega}^{2}\sum_{i=1}^{3}\overline{x_{i}}^{2}$, where the geometric frequency is defined as $\overline{\omega}=(\omega_{x}\omega_{y}\omega_{z})^{1/3}$ and $\overline{x}_{i}=x_{i}\omega_{i}/\overline{\omega}$. Upon the condition $\mu_{\mathrm{TF}}=V(\mathbf{r})$, one finds the size of the condensate confined in this potential, namely, the Thomas-Fermi radius $\overline{R}=\sqrt{\frac{2\mu_{\mathrm{TF}}}{m_{B}\overline{\omega}^{2}}}$ and the chemical potential depending on the number of particles of the condensate. Thus, from the normalization condition $N_{B}=\int d^{3}\mathbf{r}n_{\mathrm{TF}}(\mathbf{r})$,  one obtains   $\mu_{\mathrm{TF}}=\frac{\hbar\overline{\omega}}{2}\left[15N_{B}\frac{a_{B}}{a_{\mathrm{ho}}}\right]^{2/5}$. The contrast for an inhomogeneous distribution can be written as

\begin{align*}\tag{B1}
S(t)&=&\frac{1}{N_{B}}\int d^{3}\mathbf{r}n_{\mathrm{TF}}(\mathbf{r})S(\mathbf{r},t)
\nonumber\\
&=&\frac{1}{N_{B}}\int d^{3}\mathbf{r}\left[\frac{\mu_{\mathrm{TF}}-V(\mathbf{r})}{T_{B}}\right]S(\mathbf{r},t).
\label{eq:contrastdeph}
\end{align*}

The contrast $S(\mathbf{r},t)$ depends on time and density in each point of space. In other words,  the spatial dependence of the contrast is implicitly encoded in the density, and thus we substitute the homogeneous density $n\rightarrow n_{\mathrm{TF}}(\mathbf{r})$ using the local density approximation.

\bibliography{Bose-polaron}

\end{document}